\documentclass[preprint2]{aastex}
\usepackage{color}

\shorttitle{Type III Burst Pairs} \shortauthors{Tan et al.}

\begin{document}

\title{Microwave Type III Pair Bursts in Solar Flares}

\author{Baolin Tan$^1$, Hana M\'esz\'arosov\'a$^2$, Marian Karlick\'y$^2$, Guangli Huang$^3$, Chengming Tan$^1$}

\affil{$^1$Key Laboratory of Solar Activity, National Astronomical
Observatories of Chinese Academy of Sciences, Beijing 100012,
China; bltan@nao.cas.cn}

\affil{$^2$ Astronomical Institute of the Academy of Sciences of
the Czech Republic, CZ--25165 Ond\v{r}ejov, Czech Republic}

\affil{$^3$ Purple Mountain Observatory of Chinese Academy of
Sciences, Nanjing 210008, China}

\begin{abstract}

Solar microwave type III pair burst is composed of normal and
reverse-sloped (RS) burst branches with oppositely fast frequency
drifts. It is the most sensitive signature of the primary energy
release and electron accelerations in flares. This work reported
11 microwave type III pair events in 9 flares observed by radio
spectrometers in China and the Czech Republic at frequency of 0.80
- 7.60\,GHz during 1994 - 2014. These type III pairs occurred in
flare impulsive and postflare phases with separate frequency in
range of 1.08 - 3.42\,GHz and frequency gap 10 - 1700\,MHz. The
frequency drift increases with the separate frequency ($f_{x}$),
the lifetime of each burst is anti-correlated to $f_{x}$, while
the frequency gap is independent to $f_{x}$. In most events, the
normal branches are drifting obviously faster than the RS
branches. The type III pairs occurring in flare impulsive phase
have lower separate frequency, longer lifetime, wider frequency
gap, and slower frequency drift than that occurring in postflare
phase. And the latter always has strong circular polarization.
Further analysis indicates that near the flare energy-release
sites the plasma density is about $10^{10} - 10^{11}$\,cm$^{-3}$
and temperature higher than $10^{7}$ K . These results provide new
constraints to the acceleration mechanism in solar flares.
\end{abstract}


\section{Introduction}

One of the most important task of solar physics is to understand
the origin of solar energetic particles which concerns several key
problems: where is the acceleration site? what factors do dominate
the acceleration process? what signals do these particles produce?
how does the energy release? how do these particles propagate and
interact with the ambient plasmas? Solar radio type III burst is a
kind of transient event of strong emission with fast negative or
positive frequency drift. The observed type III bursts usually
occur at decimeter, meter and longer wavelengths with negative
frequency drifts (e.g., Lin \& Hudson 1971, Kane 1981, Huang et
al. 2011) which are called normal type III bursts. They are
interpreted as caused by energetic electrons streaming through the
background plasma at speed of about 0.1 - 0.9$\mathrm{c}$
($\mathrm{c}$ is the light speed) along the open coronal magnetic
field. Therefore, they are believed to be the excellent tracer of
energetic electron beams in the corona (e.g., Lin \& Hudson 1971,
Lin et al. 1981, Christe et al. 2008, Chen et al. 2013, Reid \&
Ratcliffe 2014, etc.).

Sometimes, radio type III bursts have positive frequency drifts at
decimeter or shorter wavelengths, called microwave type III bursts
in previous publications (Stahli \& Benz 1987, Sawant et al. 1994,
M\'esz\'arosov\'a et al. 2008). We call them uniformly
reverse-slope (RS) type III bursts. They are explained as caused
by energetic electron beams propagating from the accelerated site
downward to denser plasma by plasma emission(PE) or electron
cyclotron maser emission (ECME, Aschwanden \& Benz 1997, Bastian
et al. 1998). Since the RS type III bursts are mainly occurring in
decimeter and centimeter wavelengths which produced very closed to
the primary energy release site of solar flares, RS type III
bursts may reveal the intrinsic properties beneath the
acceleration site (Altyntsev et al. 2007).

Occasionally, radio type III pairs were also observed (Huang et
al. 1998, Ning et al. 2000, Ma et al., 2008, Karlick\'y 2014). A
type III pair is composed of two type III burst branches beginning
almost at the same time, one is a normal type III burst with
negative frequency drift and the other is a RS type III burst with
positive frequency drift. The normal branch can track the upgoing
energetic electron beam, and the RS branch can track the downward
electron beam (Robinson \& Benz 2000). The type III pairs can be
regarded as a sensitive tool to diagnose the physical conditions
around the flare energy-release site where magnetic reconnection
and particle acceleration takes place. For example, the separate
frequency between the normal and RS branches may pinpoint the
acceleration site and demarcate the electron density. The
frequency gap, frequency drift, and burst lifetime may reveal the
intrinsic nature of the primary energy release processes in solar
flares (Aschwanden \& Benz 1997, Sakai et al. 2005, Altyntsev et
al. 2007, Li et al. 2011).

Generally, an ideal type III pair is just composed of a normal
branch and a corresponding RS branch simultaneously (Aschwanden et
al. 1993, Huang et al. 1998, Ning et al. 2000). However, the
actual conditions are always complex, including irregular magnetic
structures, repeatedly electron acceleration, free-free
absorption, and rapid changes in the source regions (Benz et al.
1992, Meshalkina et al. 2004). Practically and generally,
observations always show a group of normal type III bursts at low
frequency band and a group of RS type III bursts at higher
frequency band occur in same time interval (Aschwanden et al.
1997, Ma et al. 2008). It is difficult to clarify their one-to-one
corresponding relationships. We call this complex assembly a type
III pair train. It may be associated to a group of upward electron
beams and a group of downward beams producing repeatedly and
propagating in the background plasmas.

Aschwanden et al. (1993) reported two radio type III pair events
with separate frequencies between the normal and RS branches at
620 and 750\,MHz, respectively. Then they collected several type
III pairs and found that separate frequency ranges in 220 -
910\,MHz with average of about 500\,MHz (Aschwanden \& Benz 1997).
They proposed that acceleration takes place in a low-density
region above a denser soft X-ray flare loop. Later, Huang et al.
(1998) reported a type III pair with separate frequency up to
1.71\,GHz. Ma et al. (2008) found that separate frequencies
scattered in a much wider range and pointed out that in different
flares the acceleration regions may locate at a large range of
heights. Are other properties of acceleration region different
from flare to flares?

This work presents a systematic investigation of solar microwave
type III pairs including events collected in the previous
literature and recently observed in two radio spectrometers
located in China and the Czech Republic. Section 2 shows the main
observed properties of type III pairs. Theoretical discussions are
in Section 3. Finally, conclusions are summarized in Section 4.

\section{Observations}

\subsection{Observation Data and Parameters}

In this work, the microwave type III pairs are observed by the
Chinese Solar Broadband Radio Spectrometers at Huairou (SBRS) and
Ond\v{r}ejov radiospectrograph in the Czech Republic (ORSC).

SBRS is a group of advanced solar radio broadband spectrometers
located in Huairou, China (Fu et al. 1995, 2004, Yan et al. 2002),
including three parts: 1.10 - 2.06\,GHz (cadence $\triangle t=5$
ms, frequency resolution $\triangle f=4$ MHz), 2.60 - 3.80\,GHz
($\triangle t=8$ ms, $\triangle f=10$ MHz), and 5.20 - 7.60\,GHz
($\triangle t=5$ ms, $\triangle f=20$ MHz). It receives total flux
of solar radio emission with dual circular polarization (left- and
right-handed circular polarization, LCP and RCP, respectively).

ORSC locates at Ond\v{r}ejov, the Czech Republic. It receives
solar radio total flux at frequencies of 0.80 - 5.00\,GHz.
$\triangle t=10$ ms. $\triangle f=5$ MHz at frequency of 0.80 -
2.00\,GHz and $\triangle f=12$ MHz at frequency of 2.00 -
5.00\,GHz (Ji\v{r}i\v{c}ka et al. 1993).

From observations of SBRS and ORSC at frequencies of 0.80 -
7.60\,GHz, we identified totally 11 microwave type III pair events
in 9 flares (Table 1). Some of them are mentioned in previous
literature (Huang et al. 1998, Ning et al. 2000, Ma et al. 2008,
Karlick\'y 2014). Here, we do not consider a few events that are
mentioned by Ma et al. (2008), because they are not clearly
visible on the observed radio spectrograms.

It is necessary to define parameters describing type III pairs.
They are including:

(1) Flare start time ($t_{fl}$), start time of the host flare
recorded at soft X-ray emission (SXR) observed by GOES.

(2) Type III time ($t_{pr}$), start time of the radio type III
pair or type III pair train.

(3) Burst lifetime ($\tau$), time interval between the start and
end of an individual type III burst.

(4) Frequency bandwidth ($f_{w}$), frequency range between the
start and end of each type III burst with emission intensity
exceeding 3 $\sigma$ above the background emission. $\sigma$ is
mean square deviation of the background emission.

(5) Frequency drift ($D$), defined as the slope of type III burst
on the spectrogram. It can be derived by using cross-correlation
of the emission profiles at each two adjacent frequency channels.
Here we make least squared Gaussian fitting curve at each
frequency channel to avoid noise fluctuations of the original
observational data. For each two adjacent frequency channels
($f_{i}$ and $f_{i+1}$), the frequency drift is:
$D_{i}=\frac{f_{i+1}-f_{i}}{t_{i+1}-t_{i}}$. $t_{i}$ and $t_{i+1}$
are times of maxima at Gaussian fitting curves at the
corresponding frequencies. For each type III burst we obtain
average frequency drift and its standard deviation ($\sigma$).
$2\sigma$ can be regarded as error bar. Fig. 1 shows an example of
a type III burst. Here, $D=1.01$ GHz s$^{-1}$, and $\sigma=0.10$
GHz s$^{-1}$. The result is expressed in $D_{r}=1.01\pm0.20$ GHz
s$^{-1}$. Using this method, frequency drifts can be measured in
the normal and RS branches of type III pairs, respectively.

\begin{figure*}[ht] 
\begin{center}
   \includegraphics[width=8.2 cm]{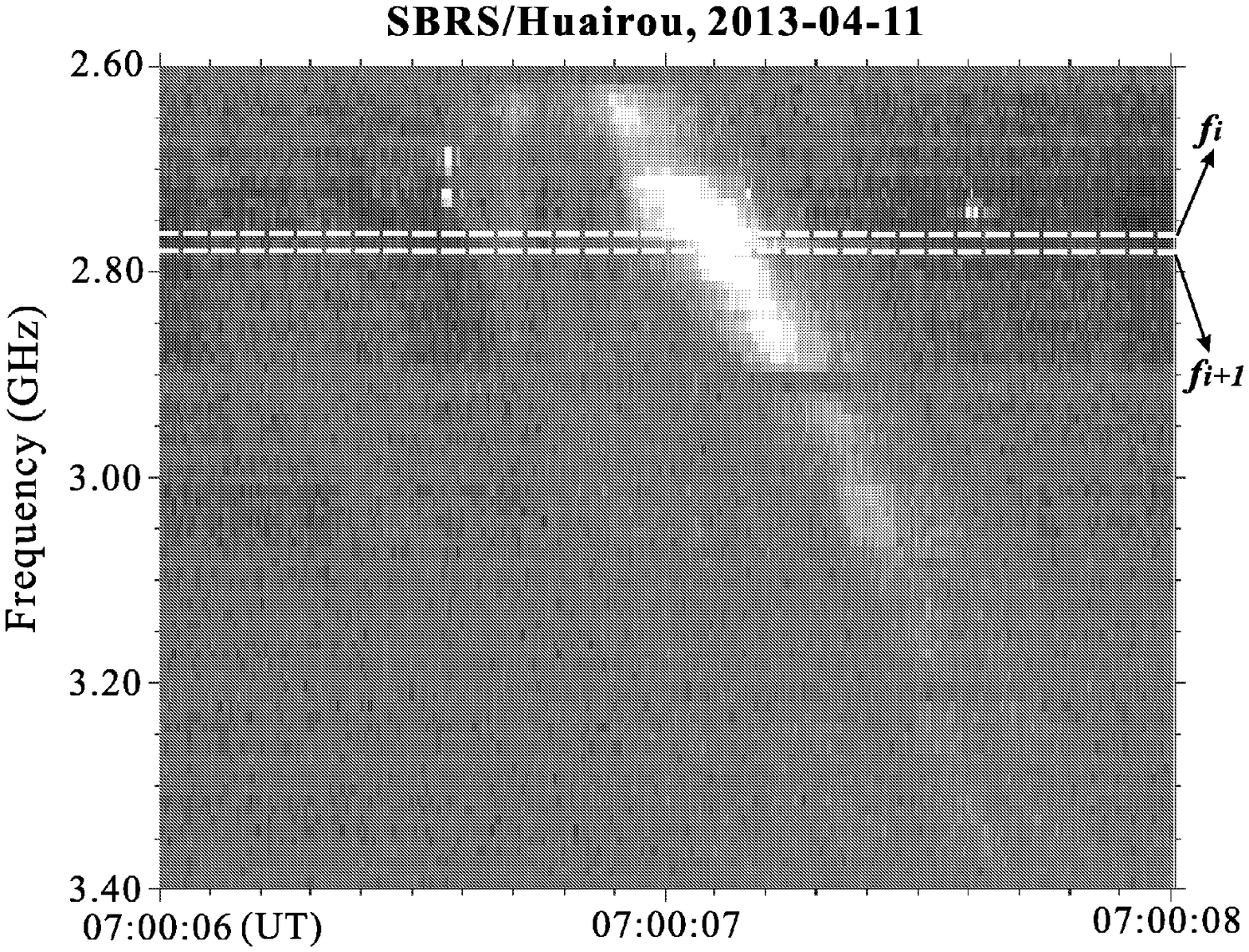}
   \includegraphics[width=8.2 cm]{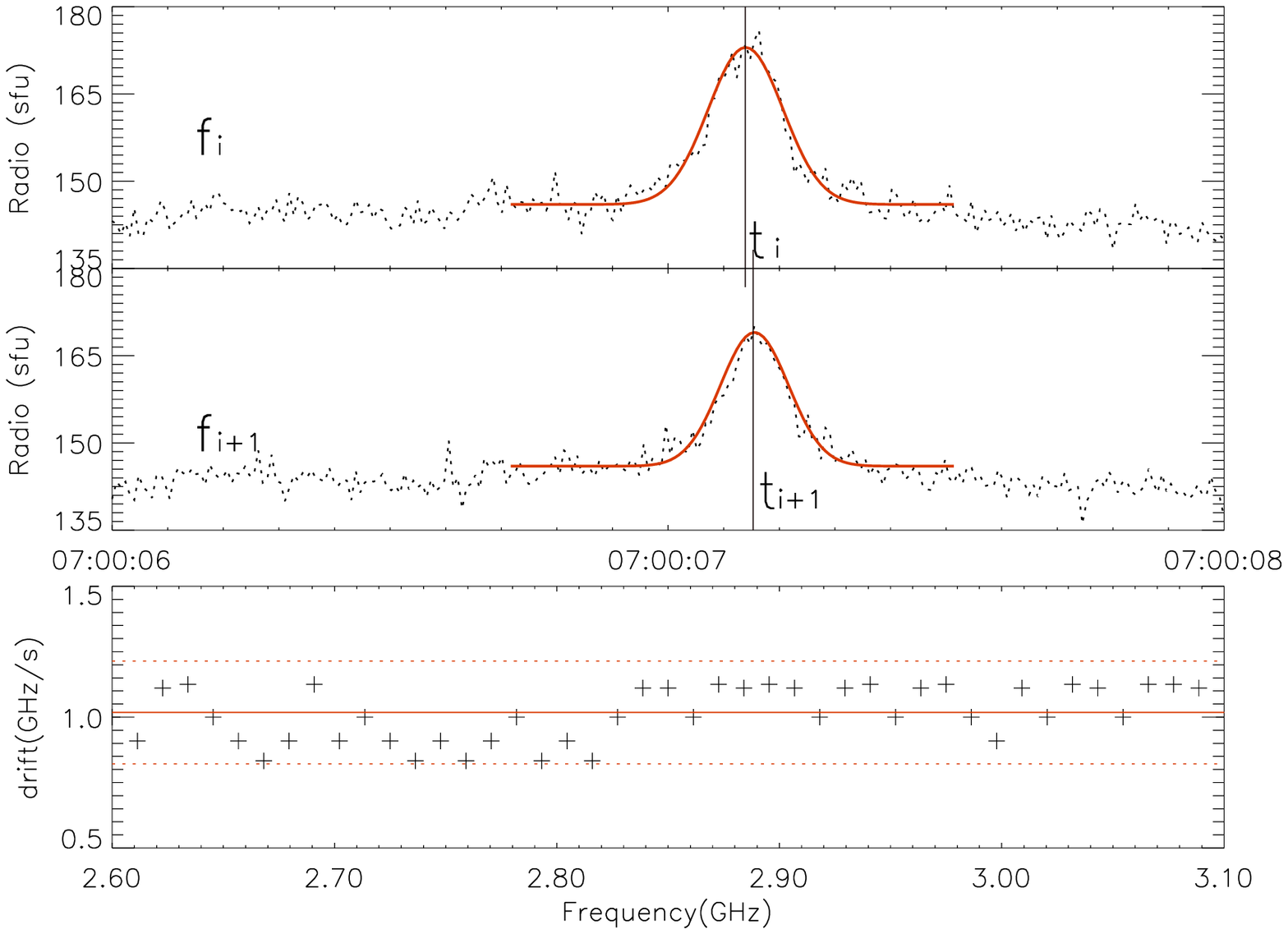}
\caption{An example of derived frequency drift rate in a type III
burst. Left panel is the spectrogram of a type III burst.
Right-upper panel shows the Gaussian fitting curves (red solid
curves) of profiles at two adjacent frequency channels.
Right-bottom is the distribution of drift rate with respect to
frequency.}
\end{center}
\end{figure*}

The relative frequency drift ($\bar{D}$) is
$\bar{D}=\frac{D}{f_{0}}$. $f_{0}$ is central frequency of the
type III burst.

(6) Separate frequency ($f_{x}$). For simple type III pair, it can
be obtained by reverse extending the normal and RS branches to a
crossing point where the frequency is $f_{x}$. Similar definition
is in Tan (2013) to describe small-scale microwave bursts. For
type III pair train, as there is no clear one-by-one relationship
between normal branches and the RS branches, $f_{x}$ can be
determined from the frequency at the watershed line (or the middle
line) between the normal and RS type III branches. Usually,
$f_{x}$ is a variable of time. We define $\frac{df_{x}}{dt}$ as
temporal change of separate frequency.

(7) Frequency gap ($\delta f$), the difference of starting
frequencies between the normal and RS branches. As for type III
pair trains, $\delta f$ is the averaged bandwidth between starting
frequencies of the normal and RS type III groups.

We also try to attribute a type III pair event with a certain
solar flare. A type III pair is related to a flare when it takes
place in the flaring duration. The flaring duration begins at the
start of a GOES SXR flare and ends when there is no obvious
decrease on the GOES SXR profile. It is possible for some events,
such as the X3.4 flare on 2006 December 13, that lasts for more
than 2 hours (Tan et al. 2010).

In following sections, several typical type III pair events are
introduced at first, and then statistical results are presented.

\subsection{Typical Type III Pair Events}

\subsubsection{Type III Pairs in the impulsive phase of a Hard X-ray Burst on 2011 September 26}

Fig. 2 presents a type III pair train at frequency of 0.80 - 2.00
GHz observed by ORSC during 06:21:30 - 06:22:00\,UT on 2011
September 26. Here, we marked 5 identified individual RS bursts
and 3 identified individual normal bursts. The normal branches
start from about 1.30\,GHz and extends below 0.80 GHz with average
frequency drift about -5.50$\pm0.49$ \,GHz\,s$^{-1}$
($\bar{D}_{n}\sim -5.0\pm0.45$ s$^{-1}$), while the RS branches
start from about 1.50 GHz and extends beyond 2.00\,GHz with
average frequency drift of 0.5$\pm0.11$\,GHz\,s$^{-1}$
($\bar{D}_{r}\sim 0.3\pm0.06$ s$^{-1}$). The middle line between
the normal and RS branches located at about 1.42 - 1.47\,GHz. That
is to say, $f_{x}$ varied from 1.42 GHz to 1.47 GHz with an
average of 1.44 GHz. $\frac{df_{x}}{dt}$ is about -4.2 MHz
s$^{-1}$. The frequency gap is about 200\,MHz. The burst lifetime
of each individual type III burst ranges from 0.14\,s to 1.10\,s
with average about 0.50\,s. The whole type III pair train lasts
for about 22\,s (Fig. 2).

\begin{figure*}[ht] 
\begin{center}
   \includegraphics[width=9.0 cm]{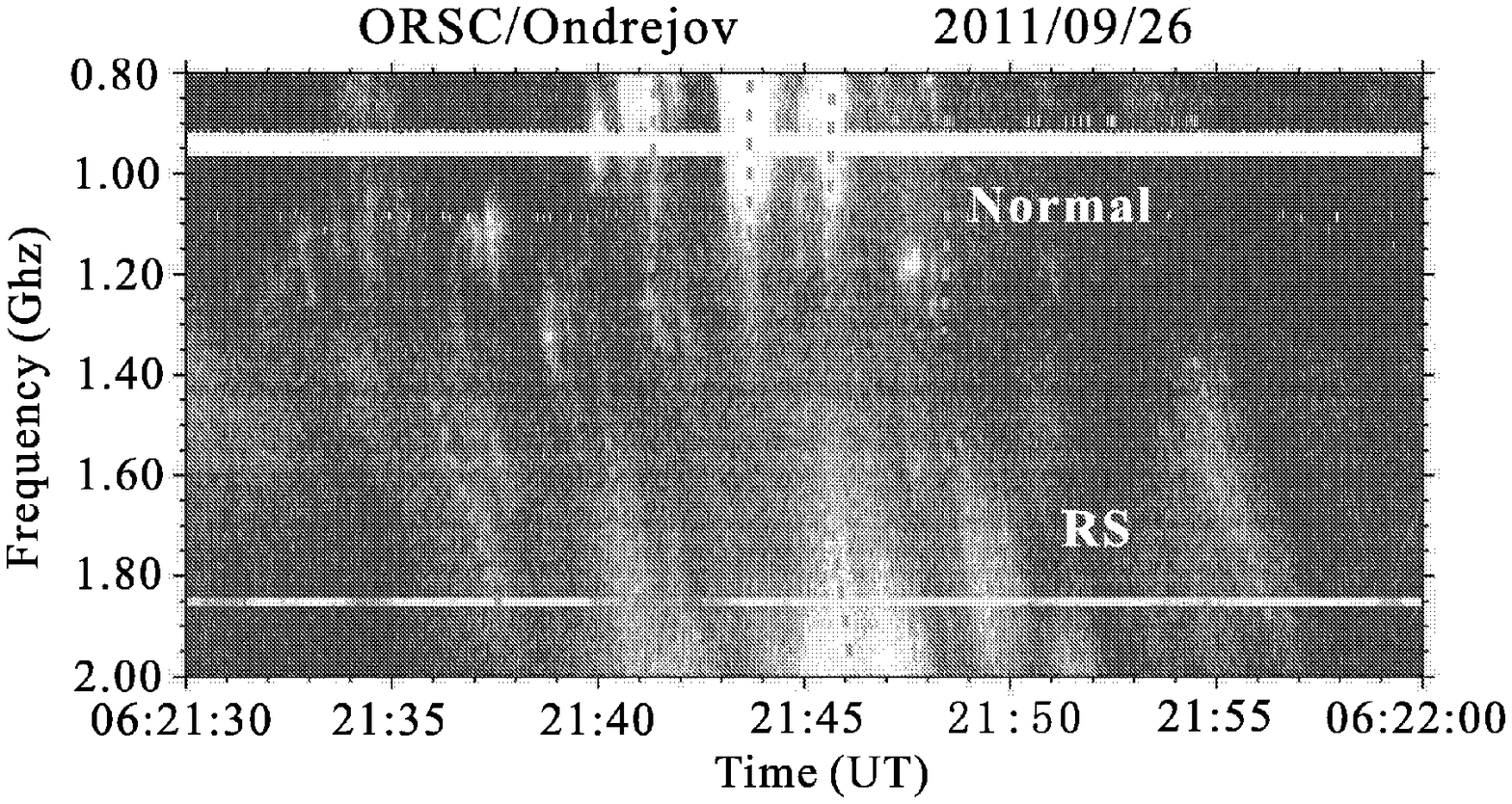}
   \includegraphics[width=7.0 cm]{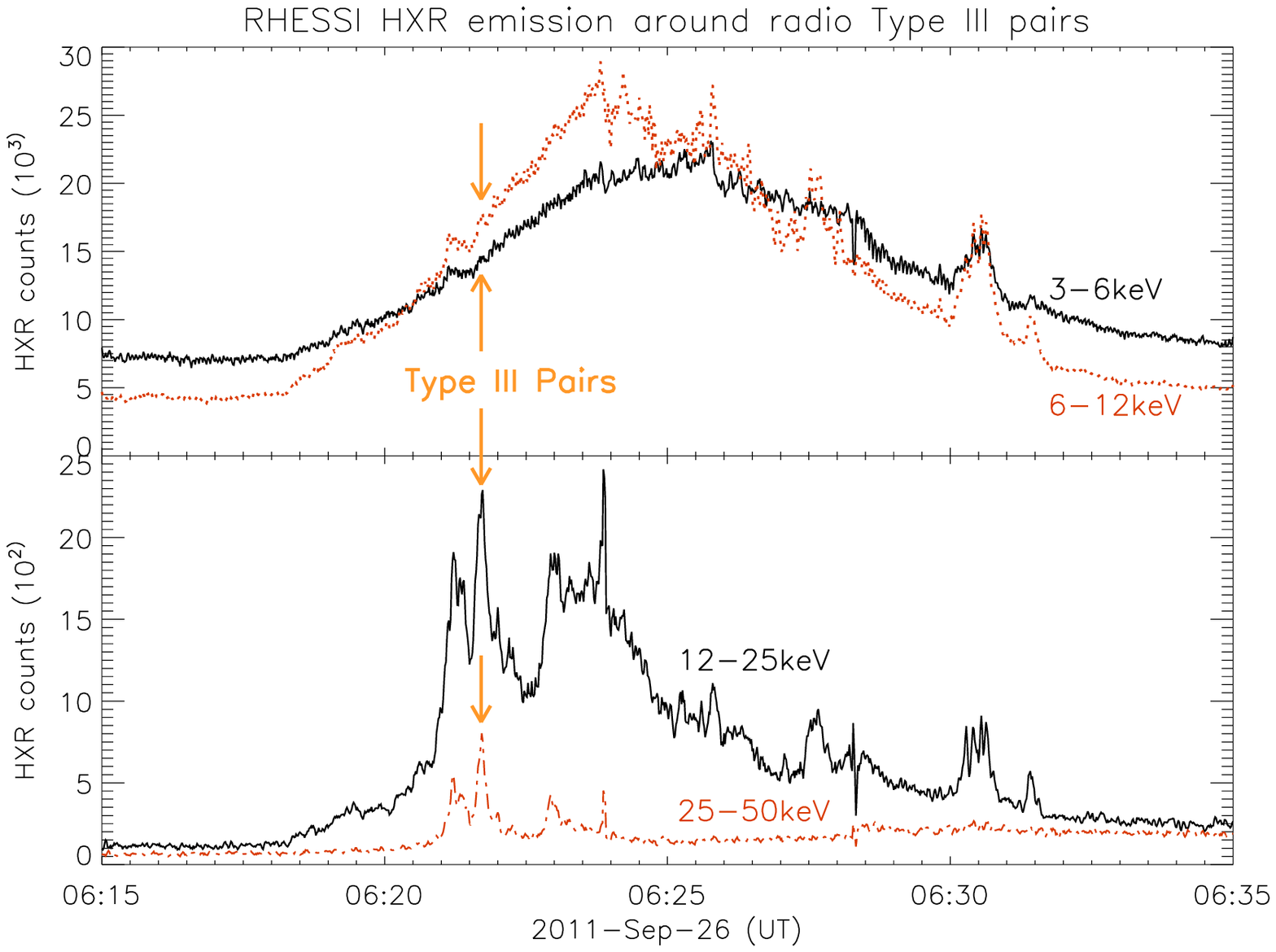}
\caption{Left panel is the spectrogram of microwave type III pair
train observed by ORSC/Ond\v{r}ejov on 2011 September 26. Green
dashed lines show ridges of type III bursts. Right panel is HXR
emission profiles observed by RHESSI where yellow arrow shows the
position of type III pairs on the HXR profile.}
\end{center}
\end{figure*}

At the same time RHESSI has perfect hard X-ray (HXR) observations.
The right panel of Fig. 2 presents HXR profiles at energies of 3 -
6\,keV, 6 - 12\,keV, 12 - 25\,keV and 25 - 50\,keV, which shows
that there is an obviously strong burst during 06:18 - 06:31\,UT
just around the above radio type III pair train. The microwave
type III pair train took place just at the flare impulsive phase
(yellow arrow in right panel of Fig. 2). The type III pair train
occur just when a strong non-thermal HXR spike occurs at energy of
12 - 25\,keV and 25 - 50\,keV. There is a strong corresponding
relation between the type III pair train and the non-thermal HXR
burst.

\subsubsection{Type III Pairs with Super Wide Frequency Gaps in a Flare on 2014 April 04}

Fig. 3 presents a type III pair train observed at frequency of
0.80 - 5.00 GHz in OSRC during 13:48:17 - 13:48:21\,UT on 2014
April 04. The left-upper panel shows the normal branches start at
about 1.60\,GHz and extend to below 0.80\,GHz with average
frequency drifts of -2.8$\pm0.80$\,GHz\,s$^{-1}$
($\bar{D}_{n}=-2.3\pm0.67$s$^{-1}$). Almost at the same time, the
RS branch starts at about 3.40\,GHz and extends to near 4.9\,GHz
with average frequency drift of 3.6$\pm1.0$\,GHz s$^{-1}$
($\bar{D}_{r}=0.9\pm0.26$s$^{-1}$). The lifetime is in a range of
0.3 - 0.7\,s with average of about 0.5\,s. The whole type III pair
train lasts for about 3\,s.

The right panel of Fig. 3 presents profiles of GOES SXR emission
at 1 - 8 \AA~ and the derived temperature in the related C8.3
flare. Here, the plasma temperature is derived from the ratio of
GOES SXR emission at wavelengths of 1 - 8\AA~ and 0.5 - 4\AA
(Thomas et al. 1985). The flare starts at 13:34\,UT, peaks at
13:48\,UT and ends at 13:56\,UT (located at N14E26 in active
region NOAA12027). The type III pairs occur at 13:43\,UT, i.e., 9
min after the flare start and 5 min before the flare maximum. This
means that the type III pairs take place in the flare impulsive
phase.

\begin{figure*}[ht] 
\begin{center}
   \includegraphics[width=8.0 cm, height=6 cm]{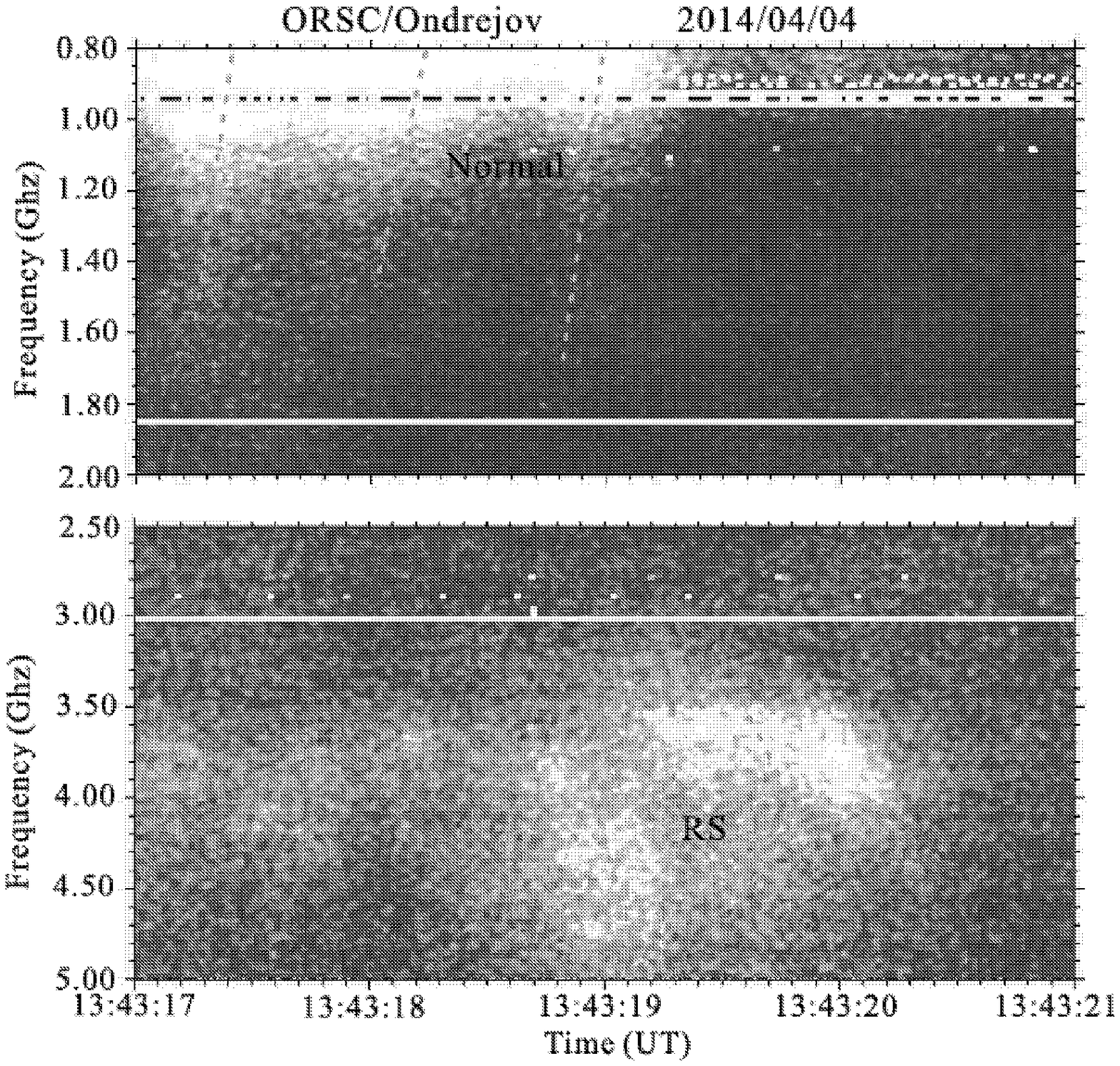}
   \includegraphics[width=8.0 cm, height=5.8 cm]{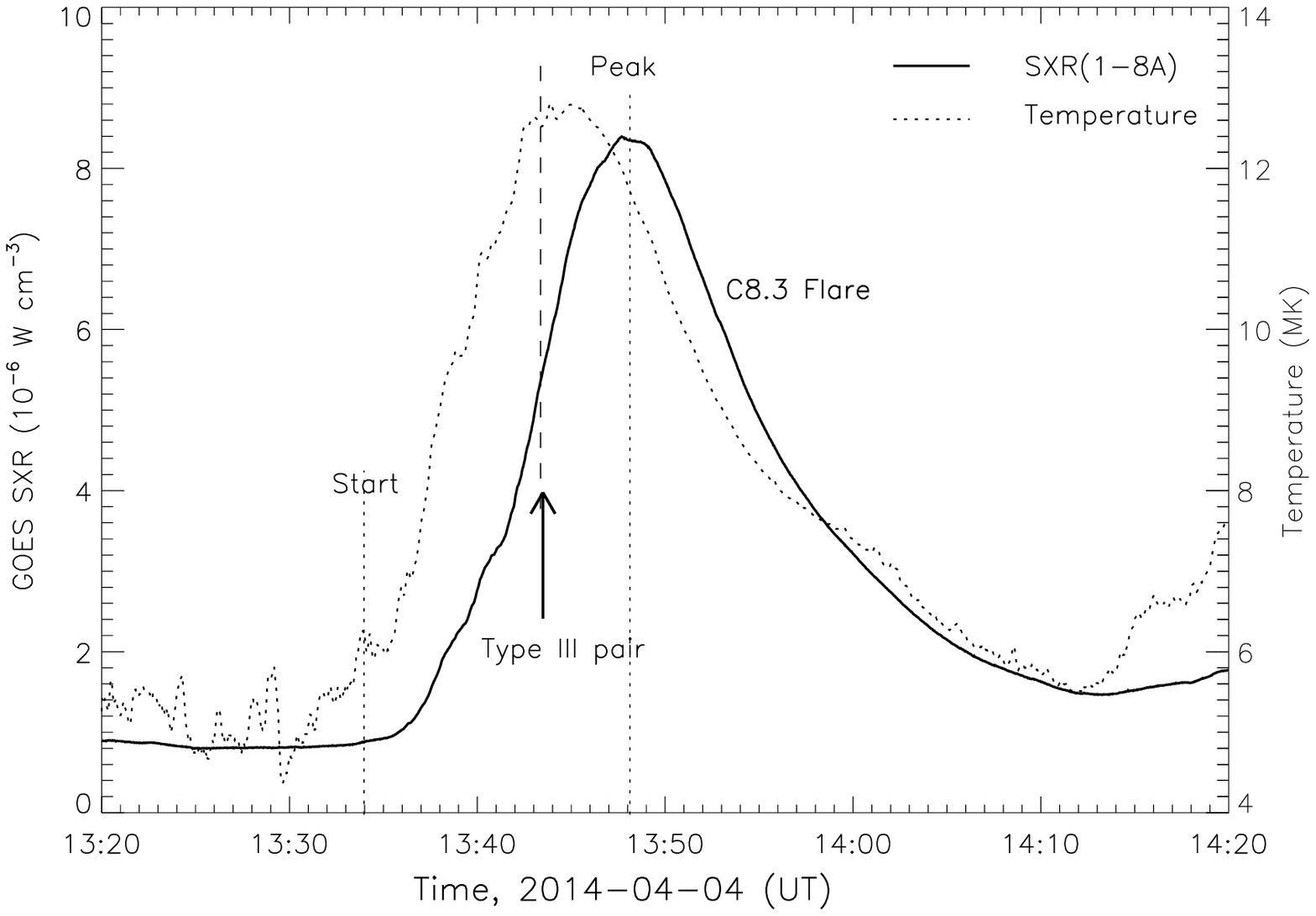}
\caption{Left panel presents spectrograms of the normal (upper)
and RS (bottom) branches of a microwave type III pairs with super
wide frequency gaps observed by OSRC on 2014 April 04. Green
dashed lines trace the ridges of type III bursts. Right panel
shows profiles of GOES SXR emission at 1-8 \AA and the temperature
in the related C8.3 flare.}
\end{center}
\end{figure*}

The most interesting of the above type III pair train is its super
wide frequency gap. Fig. 3 indicates that the frequency gap starts
from about 1.60\,GHz, ends at about 3.30\,GHz, and the gap is
about 1.70\,GHz, much wider than the other events reported in the
previous publications (Aschwanden et al. 1997, Ma et al. 2008). We
attribute this event to a type III pair event just because the
normal and RS branches took place almost simultaneously in a short
period (about 4 s) at the early impulsive phase of the C8.3 flare.
On the other hand, we have scrutinized the multi-wavelength EUV
images of AIA/SDO and found that there was no clear evidence to
show any bursts occurring in different loops in above duration
which ruled out the possibility that the above normal and RS type
III bursts might generate in different acceleration sites from
different coronal loops.

The temperature profile shows the above type III pairs occurred
just at the time of maximum temperature which implies that the
maximum temperature may play an important role in the formation of
the super wide frequency gap.

\subsubsection{Several Type III Pair Events in Postflare Phase on 2006 December 13}

Sometimes, microwave type III pairs occurred in the postflare
phase. Fig. 4 presents three microwave type III pair trains at
frequency of 2.60 - 3.80 GHz in the postflare phase of a
long-duration powerful X3.4 flare on 2006 December 13. All of them
are strongly right-handed circular polarization overlapping on a
long-duration broadband type IV continuum. The flare starts at
02:14\,UT, peaks at 02:40\,UT and ends at 02:57\,UT at SXR
emission. The microwave burst lasts to even after 04:45 UT (see
right-bottom panel of Fig. 4). In fact, the flare active region
NOAA 10930 is an isolated one on solar disk accompanying microwave
bursts with many kinds of fine structures, such as quasi-periodic
pulsations, spikes, and type III bursts, etc. (Tan et al. 2007,
2010).

\begin{figure*}[ht] 
\begin{center}
   \includegraphics[width=8.2 cm, height=6.0cm]{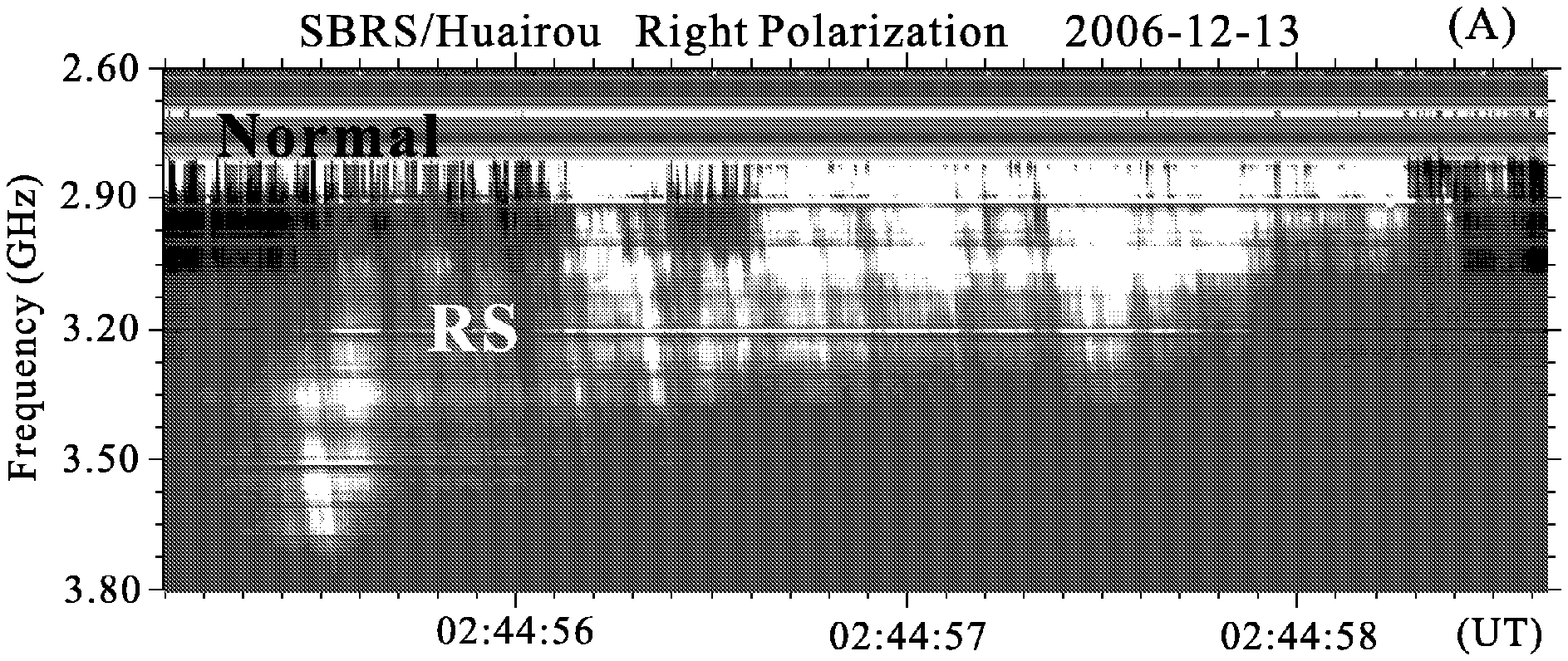}
   \includegraphics[width=8.2 cm, height=6.0cm]{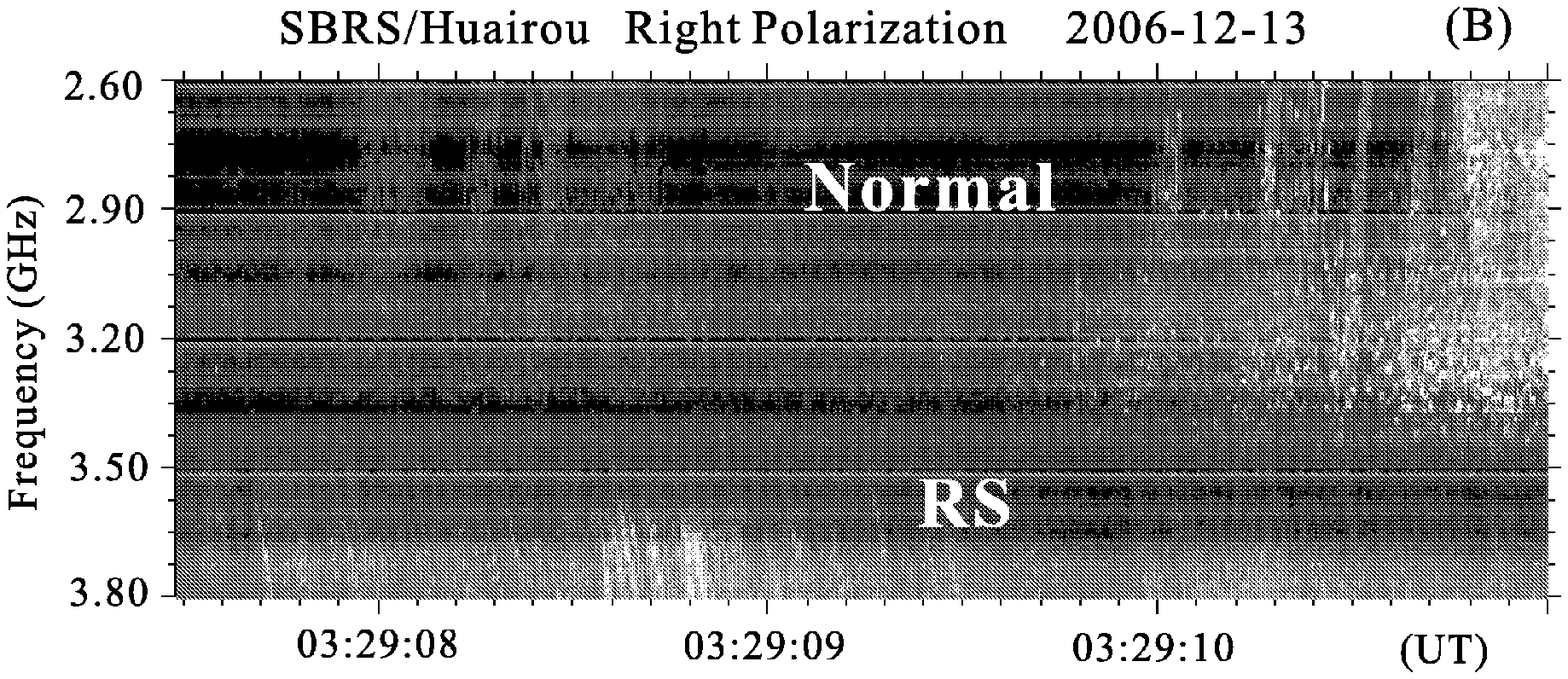}
   \includegraphics[width=8.2 cm, height=6.0cm]{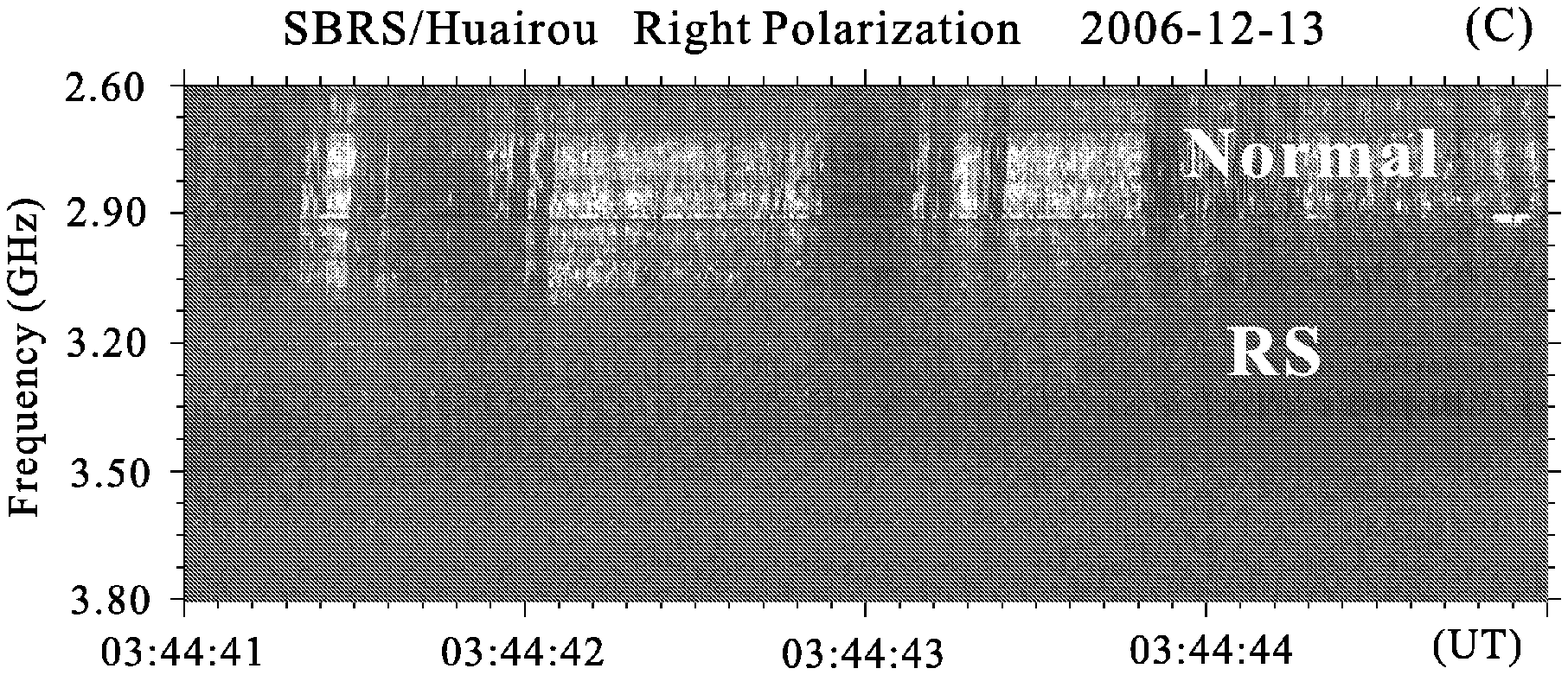}
   \includegraphics[width=8.2 cm, height=5.5cm]{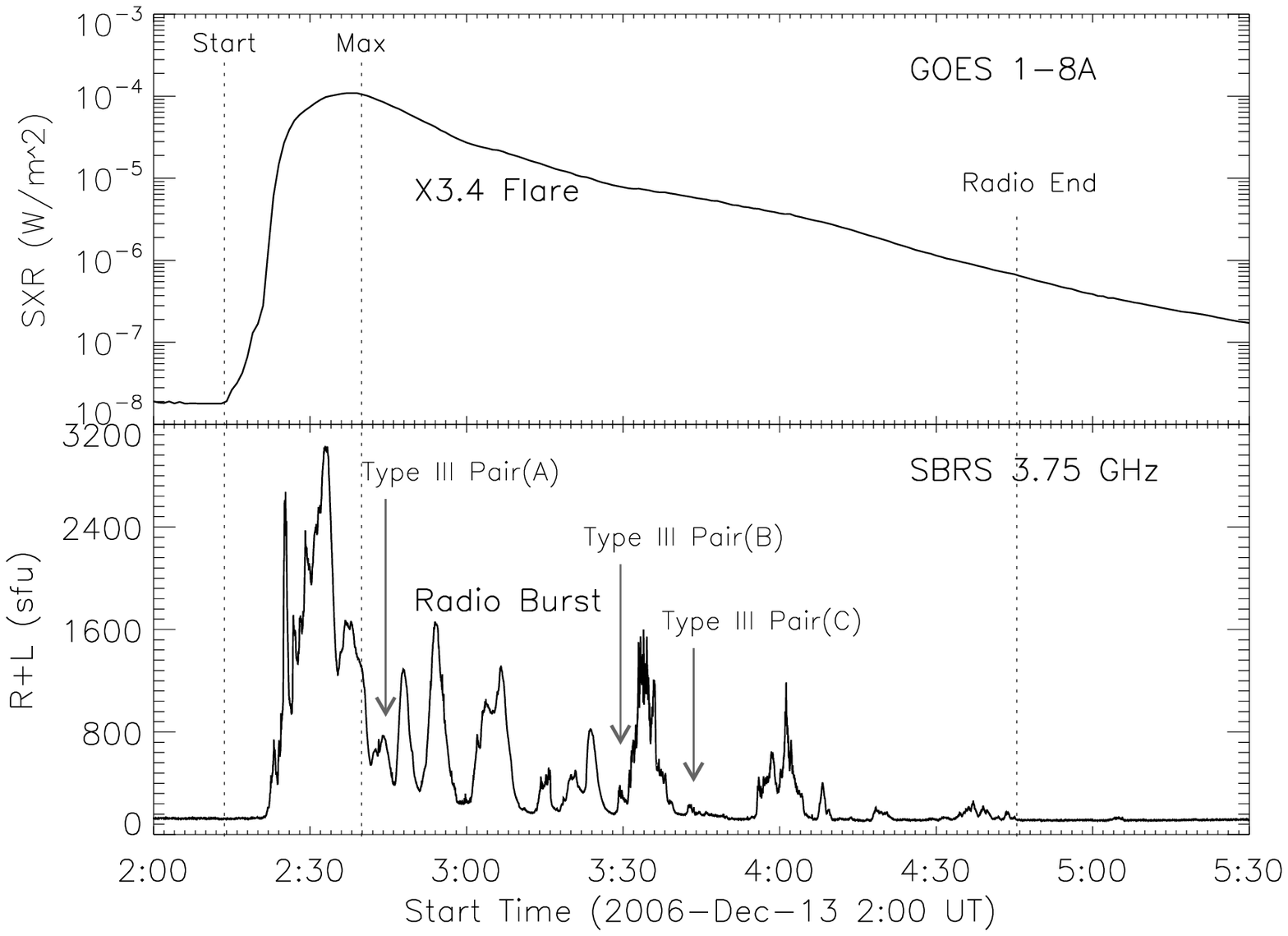}
\caption{Three microwave type III pair trains at 02:44:56\,UT (A),
03:29:09\,UT (B), and 03:44:43\,UT (C) observed at SBRS/Huairou on
2006 December 13. The right-bottom panel shows the whole flare
process profiled by GOES SXR at 1-8 \AA~ and microwave emission at
3.75 GHz.}
\end{center}
\end{figure*}

Panel (A) of Fig. 4 shows a type III pair train occurring at 02:45
UT, near the flare maximum. The normal branches start from
2.90\,GHz and extend to below 2.60\,GHz with an extremely fast
frequency drift of -14.5$\pm$0.6\,GHz\,s$^{-1}$. The burst
lifetime is in a range of 140 - 180\,ms with average of about
160\,ms. The RS branches start from 2.95 - 2.92\,GHz and extend to
3.3 - 3.6\,GHz with average frequency drift of about 8.6$\pm$0.8
GHz s$^{-1}$, the burst lifetime is 0.1 - 0.3\,s with average of
about 0.2\,s. The whole type III pair train lasts for about 3\,s.
The separate frequency decreases slowly from 2.93\,GHz to
2.91\,GHz, $\frac{df_{x}}{dt}\sim$-5 MHz s$^{-1}$, and the
frequency gap is about 50\,MHz.

Panel (B) shows another type III pair train occurring around
03:29\,UT, about 49 min after the flare maximum. The normal
branches start at about 3.30 - 3.40\,GHz and extend to below
2.60\,GHz with negative frequency drift of about
-13.1$\pm$0.80\,GHz\,s$^{-1}$. The burst lifetime is in a range of
60 - 120\,ms with average of about 100\,ms. The RS branches start
at about 3.50 - 3.60\,GHz and extend beyond 3.80\,GHz with
$\bar{D}\approx$1.5$\pm$0.16\,s$^{-1}$, the burst lifetime is 40 -
60\,ms with average of about 50\,ms. This train lasts for about
3.5\,s. The separate frequency increases slowly from 3.40\,GHz to
3.44\,GHz, $\frac{df_{x}}{dt}\sim$10 MHz s$^{-1}$, and the
frequency gap is about 260\,MHz.

Panel (C) shows a type III pair train occurring at about 03:45 UT,
far from the flare maximum and still in the postflare decay phase.
Its separate frequency is about 2.90\,GHz and with a very small
frequency gap of about 10\,MHz. The burst lifetime of the normal
branches is 50 - 80\,ms (averaged 70\,ms). The event lasts for
about 4\,s. The relative frequency drifts of the normal and RS
branches are -3.8$\pm$0.44\,s$^{-1}$ and 3.2$\pm$0.26\,s$^{-1}$,
respectively. The separate frequency is nearly a constant of
2.90\,GHz during the type III pairs. This train behaves as
absorption structure which is analyzed by Chen \& Yan (2008).

The right-bottom panel of Fig. 4 presents the comparison between
SXR intensities and the microwave emission at frequency of 3.75
GHz. Each train took place just at a short microwave burst showed
as arrows. Therefore, we proposed that the above three type III
pair trains were related to the X3.4 flare. At the same time, we
also found that most RS branches emission are relatively weaker
than the corresponding normal branches in the above three type III
pair trains.

\subsection{Statistic Properties}

Table 1 and Table 2 list 11 microwave type III pair events. Among
them, three events (19940105, 20110215, and 20140404) composed of
single normal branch and single RS branch. The other 8 events are
type III pair trains composing of normal groups and RS burst
groups. We make a statistical analysis in this section.

Browsing active regions of the host flares related to the type III
pairs we find that all the flares tend to locate near the center
of solar disk. For example, the X3.4 flare associated with several
type III pair events on 2006 December 13 is located in S06W21. All
other host flare active regions are located from east 35$^{\circ}$
to west 25$^{\circ}$ and from north 30$^{\circ}$ to south
21$^{\circ}$ on the solar disk. The host flares include from
extremely powerful (X-class) to rather weak ones (C-class) and
from long-duration flares (e.g., X3.4 flare on 2006-12-13 lasts
for more than 2 hours) to very short duration flares (e.g., C8.3
flare in 2014-08-01 lasts for only 10 min).

\begin{deluxetable}{ccccccccccccccccc}
\tablecolumns{16} \tabletypesize{\scriptsize} \tablewidth{0pc}
\tablecaption{Properties of the microwave type III pairs occurring
in flare impulsive phases\label{tbl-1}} \tablehead{
 \colhead{Event}             &\colhead{1994-01-05}  &\colhead{1998-04-15}  &\colhead{2012-02-12}  &\colhead{2011-09-26} &\colhead{2012-07-02}  &\colhead{2014-04-04}   &\colhead{2014-08-01}   \\
}
  \startdata
  Flare Class                      &     M1.0(a)    &       C8(a)    &   M8.3(b)      &  HXR burst(b) &   M5.6(b)       &  C8.3(b)        &   M2.0(b)      \\
  Location                         &     ---        &     N30W25     &  N21E07        &  N12E22       &   S17E06        &  N14E26         &   S09E35       \\
  $t_{fl}$ (UT)                    &    06:45       &      07:37     &   11:19        &  06:18        &   10:43         &  13:34          &   14:43        \\
  $t_{mx}$ (UT)                    &    07:02       &      07:47     &   11:27        &  06:25        &   10:52         &  13:48          &   14:49        \\
  $t_{pr}$ (UT)                    &    06:52       &      07:42     &   11:24        &  06:21        &   10:50         &  13:43          &   14:47        \\
  $\tau$ (s)                       &    1.60        &      0.80      &   0.15         &  0.50         &   0.50          &  0.50           &   1.10         \\
  Pol                              &     weak       &      weak      &  (strong)      &  (weak)       &   (weak)        &  (weak)         &   (weak)       \\\hline
  $f_{n0}$ (GHz)                   &     1.25       &      1.62      &   0.99         &  1.10         &   2.35          &  1.20           &   1.03         \\
  $f_{wn}$ (MHz)                   &     500        &      120       &   180          &  $>$600       &   700           &  $>$800         &   350          \\
  $D_{n}$ (GHz/s)                  & -0.7$\pm$0.2   &  -0.8$\pm$0.2  & -0.6$\pm$0.1   & -5.5$\pm$0.49 &  -1.6$\pm$0.3   &  -2.8$\pm$0.8   & -0.4$\pm$0.2   \\
  $\bar{D}_{n}$ ($s^{-1}$)         & -0.6$\pm$0.16  &  -0.5$\pm$0.12 & -0.6$\pm$0.1   & -5.0$\pm$0.45 &  -0.7$\pm$0.13  &  -2.3$\pm$0.67  & -0.4$\pm$0.2   \\\hline
  $f_{r0}$ (GHz)                   &     1.85       &    1.84        &   1.15         &  1.75         &   4.15          &  3.85           &   1.85         \\
  $f_{wr}$ (MHz)                   &    $>$300      &    120         &   100          &  500          &   1500          &  700            &   $>$300       \\
  $D_{r}$ (GHz/s)                  &  0.5$\pm$0.2   &  0.6$\pm$0.2   &  0.5$\pm$0.1   & 0.5$\pm$0.11  &  2.8$\pm$0.4    &  3.6$\pm$1.0    & 0.6$\pm$0.2    \\
  $\bar{D}_{r}$ ($s^{-1}$)         &  0.3$\pm$0.11  &  0.3$\pm$0.16  &  0.4$\pm$0.08  & 0.3$\pm$0.06  &  0.7$\pm$0.1    &  0.9$\pm$0.26   & 0.3$\pm$0.10   \\\hline
  $f_{x}$ (GHz)                    &     1.71       &    1.73        &   1.08         &  1.44         &   3.00          &  2.40           &   1.45         \\
  $df_{x}/dt$ (MHz/s)              &    ---         &    3.0         &   -3.5         &  4.2          &   -9.0          &   ---           &   $\sim$ 0     \\
  $\delta f$ (MHz)                 &     140        &    10          &   35           &  200          &   1000          &  1700           &   500          \\
  $T$($\times10^{7}$ K)            &    1.08        &    1.12        &   1.51         &  1.60         &   1.79          &  1.26           & 1.27           \\
  $n_{x}$($\times10^{16}$m$^{-3}$) &    0.90        &    0.92        &   1.4          &  0.64         &   2.8           &  1.8            &     0.65       \\
   \enddata
 \tablecomments{$a$: observed by SBRS, $b$: observed by ORSC, $t_{fl}$: flare start time, $t_{mx}$: the time of flare maximum, $t_{pr}$: type III time, $\tau$: burst lifetime, Pol: circular
  polarization, $f_{n0}$ and $f_{r0}$: central frequency, $f_{wn}$ and $f_{wr}$: frequency bandwidth, $D_{n}$ and $D_{r}$: frequency drifts, $\bar{D}_{n}$ and
  $\bar{D}_{r}$: relative frequency drifts, $f_{x}$: separate frequency, $df_{x}/dt$: temporal change of separate frequency, $\delta f$: frequency gap, $T$: plasma temperature, $n_{x}$: plasma density.}
\end{deluxetable}

\begin{deluxetable}{ccccccccccccccccc}
\tablecolumns{16} \tabletypesize{\scriptsize} \tablewidth{0pc}
\tablecaption{Properties of the microwave type III pairs occurring
in postflare phases \label{tbl-1}} \tablehead{
 \colhead{Event}                    &\colhead{2006-12-13A}    &\colhead{2006-12-13B}   &\colhead{2006-12-13C}  &\colhead{2011-02-15}  \\
}
  \startdata
  Flare Class                       &    X3.4(a)       & X3.4(a)         &   X3.4(a)       &  X2.2(a)        \\
  Location                          &    S06W21        & S06W21          &   S06W21        &  S20W10         \\
  $t_{fl}$ (UT)                     &    02:14         & 02:14           &   02:14         &  01:46          \\
  $t_{mx}$ (UT)                     &    02:40         & 02:40           &   02:40         &  01:56          \\
  $t_{pr}$ (UT)                     &    02:45         & 03:29           &   03:45         &  02:03          \\
  $\tau$ (s)                        &    0.16          & 0.10            &   0.07          &  0.25           \\
  Pol                               &    strong        & strong          &   strong        &  strong         \\\hline
  $f_{n0}$ (GHz)                    &     2.75         & 2.98            &   2.75          &  2.80           \\
  $f_{wn}$ (MHz)                    &   $>$300         & $>$750          &   $>$300        &  200            \\
  $D_{n}$ (GHz/s)                   &  -14.5$\pm$0.6   & -13.1$\pm$0.8   &  -10.5$\pm$1.2  & -0.9$\pm$0.2    \\
  $\bar{D}_{n}$ ($s^{-1}$)          &  -5.3$\pm$0.22   & -4.4$\pm$0.29   &  -3.8$\pm$0.44  & -0.3$\pm$0.07   \\\hline
  $f_{r0}$ (GHz)                    &     3.28         & 3.68            &   3.00          &  3.00           \\
  $f_{wr}$ (MHz)                    &     650          & $>$250          &   200           &  100            \\
  $D_{r}$ (GHz/s)                   &  8.6$\pm$0.8     & 5.6$\pm$0.6     &  9.7$\pm$0.8    & 1.5$\pm$0.4     \\
  $\bar{D}_{r}$ ($s^{-1}$)          &  2.6$\pm$0.24    & 1.5$\pm$0.16    &  3.2$\pm$0.27   & 0.5$\pm$0.14    \\\hline
  $f_{x}$ (GHz)                     &    2.92          & 3.42            &   2.90          &  2.93           \\
  $df_{x}/dt$ (MHz/s)               &    -5.0          & 10.0            &   $\sim$ 0      &  ---            \\
  $\delta f$ (MHz)                  &    50            & 260             &   10            &  30             \\
  $T$ ($\times10^{7}$K)             &   1.85           & 1.45            &   1.38          & 1.68            \\
  $n_{x}$ ($\times10^{16}$m$^{-3}$) &   10.5           &     14.4        &    10.3         &  10.6           \\
   \enddata
 \tablecomments{same as Table 1}
\end{deluxetable}

The most important parameter of type III pairs is the separate
frequency which is related to the height and plasma density of the
acceleration region. However, it is different from flare to flares
because of different magnetic field conditions, energy release
mechanism, and particle acceleration processes. These processes
are possibly related to the burst lifetime, frequency drift, and
frequency gap. Therefore, the correlations between the separate
frequency and the other parameters may reveal some regular
variations. Fig. 5 presents these correlations where dotted lines
are created by least-squared linear fitting. The correlate
coefficients are marked beside the fitting lines. From the
analysis, we found following regular variations:

(1) All microwave type III pairs occurred in impulsive and
postflare phase after the flare onset. Aschwanden \& Benz (1995)
proposed that the particle acceleration might occur from the
preflare phase, and therefore the type III pairs can be observed
in this phase. However, our work shows that no type III pair
occurred in preflare phase, but some microwave type III pairs
appeared in the postflare phase.

(2) The parameters of type III pairs distribute in a large range.
The lifetime is 0.07 - 1.60 s, separate frequency is 1.08 -
3.42\,GHz which is much higher and wider than the previous
reported results (Aschwanden et al. 1997, Ning et al. 2000, etc.),
and frequency gap is 10\,MHz - 1.70\,GHz, frequency drift is 0.4 -
14.5\,GHz\,s$^{-1}$ and the corresponding relative frequency drift
is 0.3 - 5.3 \,s$^{-1}$.

(3) The peculiar property is that normal branches drift faster
than the RS branches in most type III pairs. The average relative
frequency drift is -1.78\,s$^{-1}$ (faster) for normal branches
and 1.04\,s$^{-1}$ (slower) for RS branches. Only two among the 11
events have faster drifting RS branches than the normal ones.

(4) Although the separate frequency distributes in a large rage,
its temporal variation is very slow in each type III pair train.
$\frac{df_{x}}{dt}$ is in a small range from -9.0 MHz s$^{-1}$ to
10 MHz s$^{-1}$. Sometimes it is even close to zero ($f_{x}$
closes to a constant, e.g., the last event in the X3.4 flare on
2006 Dec 13).

(5) The burst lifetime is anti-correlated obviously to the
separate frequency at confidence level $>90\%$ (panel 1).

(6) Frequency drifts increase with the separate frequencies. The
frequency drift of normal branches is positively related to
separate frequency at confidence level $>95\%$ (panel 2) while RS
branches also have a similar correlation at a lower confidence
level $>90\%$ (panel 3).

(7) Frequency gap is independent to the separate frequency (panel
4).

(8) The type III pair events occurring in flare impulsive phase
are different obviously than those occurring in the postflare
phase. Type III pair events in flare impulsive phase (Table 1)
have separate frequencies at 1.08 - 3.00\,GHz (average 1.83\,GHz),
burst lifetime 0.15 - 1.60\,s (average 0.81\,s), large frequency
gaps 10 - 1700\,MHz (average 512\,MHz), and slow relative
frequency drifts (average 0.83\,s$^{-1}$ for normal and 0.51
\,s$^{-1}$ for RS branches, respectively). All type III pairs with
super wide frequency gaps ($\geq500$ MHz) are occurring in the
flare impulsive phase and very close to the flare maximum. Fig. 3
shows that the type III pair with super wide frequency gap occurs
near the flare maximum temperature. The other type III pairs with
frequency gap of $>$500\,MHz also occurred very close to the
maximum temperature. This fact implies that the state of
temperature may play important role in the formation of type III
pairs. As a comparison, type III pairs in postflare phases (Table
2) are related to some powerful flares (X-class). They have higher
separate frequencies at 2.90 - 3.42\,GHz (average 3.04\,GHz),
smaller frequency gaps 10 - 260\,MHz (average 87.5\,MHz), very
short lifetime 0.07 - 0.25\,s (average 0.15\,s), and fast relative
frequency drifts (average 3.45\,s$^{-1}$ for normal branches and
1.95 \,s$^{-1}$ for RS branches, respectively). All the four
events have strong circular polarization.

\begin{figure}[ht] 
\begin{center}
     \includegraphics[width=8.2 cm]{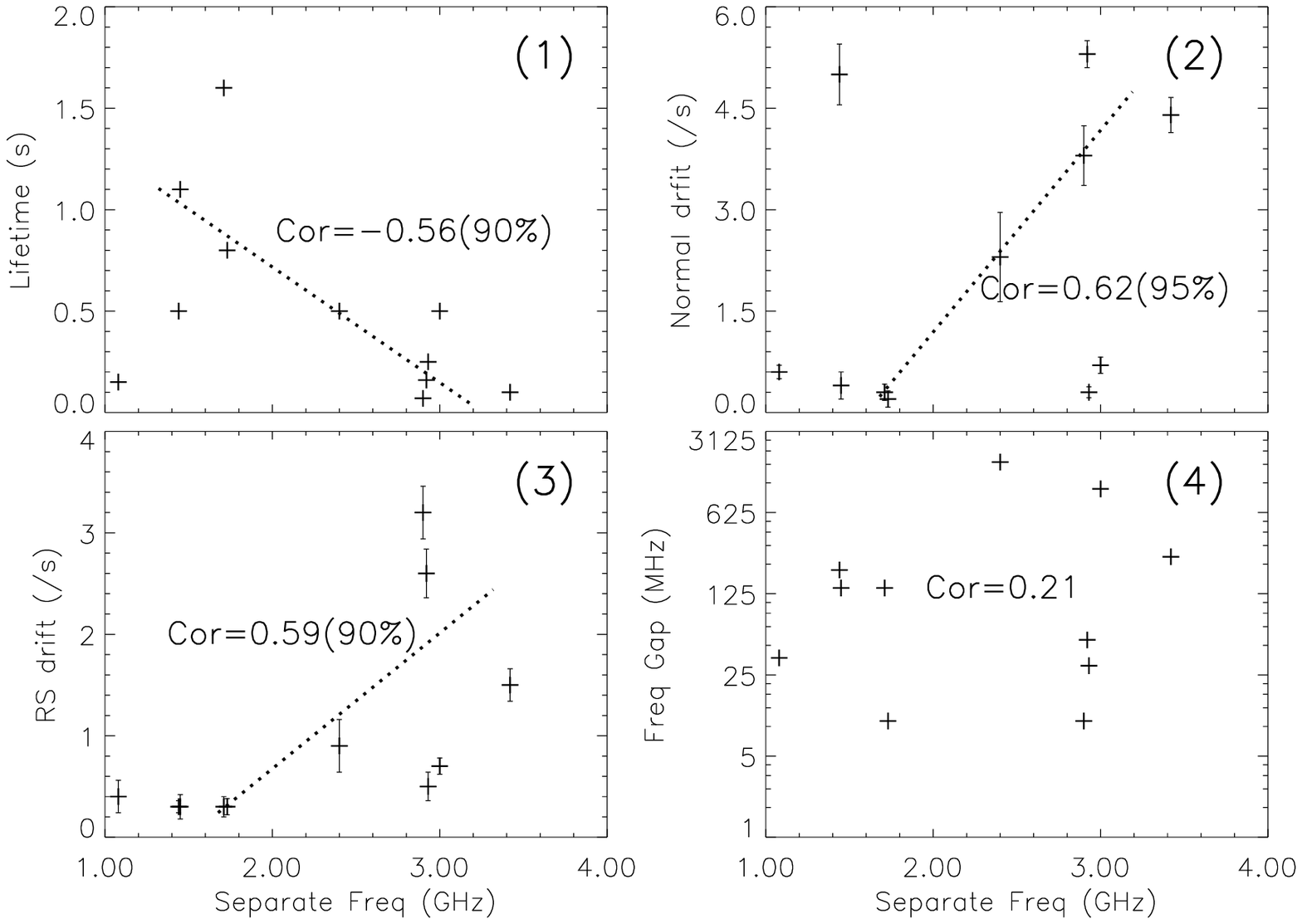}
\caption{Distributions of burst lifetime, frequency drift, and
frequency gap with respect to the separate frequency. The dotted
lines are linear fittings. $Cor$ is the correlation coefficient
with confidence level in brackets. Error bar is scaled in two
times of standard deviation ($2\sigma$).}
\end{center}
\end{figure}

Among the 11 events, there are observations of polarization in 6
event obtained by SBRS, while the other 5 OSRC-observed events
have no observations of polarization. Here we use a similarity of
the OSRC-observed events and SBRS-observed events to deduce their
polarizations. Because the observed type III pairs here with known
weak polarization have long lifetimes ($\tau>$ 0.50 s), while the
type III pairs with known strong polarization have short lifetimes
($\tau<$ 0.3 s). When $\tau<$ 0.5 s we suppose its polarization is
strong, and when $\tau\geq$ 0.5 s its polarization is assumed to
be weak. Table 1 lists the status of polarization in each type III
pair event. We find that the type III pairs in postflare phase are
strong circular polarizations. Among the 7 type III pair events in
impulsive phase, only one event has strong polarization
(20120212).

However, as our statistical sample only contains 11 events, the
above results are rather approximative. They are scattering in a
wide range. Therefore, the results mentioned here just provide a
preliminary sketchy statistical regularities.

\section{Theoretical Discussion}

Generally, it is believed that microwave type III pairs are
signatures of bi-directional energetic electron beams which may
pinpoint acceleration sites in solar flares. The magnetic
reconnection may take place in a cusp configuration above and very
close to the top of flaring loop similarly as the Masuda-like
flare (Masuda et al. 1994, etc.), or in the current sheet above
and beyond flare loops triggered by the tearing-mode instability
(Kliem et al. 2000, etc). Electrons are accelerated in the
reconnection site, propagate upward or downward, and produce
normal or RS type III bursts. Therefore, the observations of
microwave type III pairs may provide the most fundamental
information of the flare energy-release region where the magnetic
reconnection, particle acceleration, and the primary energy
release occur.

The short lifetimes and very high brightness temperatures of
microwave type III bursts indicate that the emission should be
coherent processes. The first possible mechanism is electron
cyclotron maser emission (ECM, Melrose \& Dulk 1982, Tang et al.
2012) which requires a strong background magnetic field:
$f_{ce}>f_{pe}$. $f_{ce}$ is electron gyro-frequency and $f_{pe}$
is plasma frequency. As the type III pairs here occurred at
frequency of 0.9 - 4.5\,GHz, which requires magnetic fields
$B>320-1600$\,G, generally too strong to occur in the flare
energy-release site. The second possible mechanism is plasma
emission (PE, Zheleznyakov \& Zlotnik 1975) which is generated
from the coupling of two excited plasma waves at frequency of
2$f_{pe}$ (second harmonic PE), or the coupling of an excited
plasma wave and a low-frequency electrostatic wave at frequency of
about $f_{pe}$ (fundamental PE). As PE has no strong magnetic
field constraints, it is much easy to be the favorite mechanism of
microwave type III bursts.

In PE mechanism, the emission frequency can be expressed,

\begin{equation}
f=\frac{s}{2\pi}(\frac{e^{2}n_{e}}{\varepsilon_{0}m_{e}})^{1/2}\approx
9sn_{e}^{1/2}.
\end{equation}
$s$ is the harmonic number, $s=1$ is the fundamental PE while
$s=2$ is the second harmonic PE. Generally, the fundamental PE has
strong circular polarization, while the second harmonic PE always
has relatively weak circular polarization. Because the separate
frequency ($f_{x}$) of the microwave type III pairs is possibly
related to the plasma emission near the acceleration site, here
the plasma density can be estimated from Equation (1):

\begin{equation}
n_{x}=f_{x}^{2}/81s^{2} ~(m^{-3}).
\end{equation}
The results are listed in Table 1 showing densities in range of
(0.64 - 2.8)$\times10^{16}$ m$^{-3}$ in impulsive phase and (1.03
- 1.44)$\times10^{17}$ m$^{-3}$ in the postflare phase,
respectively. These densities are about one order of magnitude
higher than that deduced by Aschwanden \& Benz (1997). This fact
indicates that the acceleration possibly takes place in the lower
corona and upper chromosphere.

The plasma emission is generated when electron beams fly off the
acceleration site and propagate upward or downward in the ambient
plasma. However, because of Coulomb collisions between the
electron beam and ambient plasma, an electron beam with velocity
of $v_{b}$ producing a microwave burst should have lifetime $\tau$
(Benz et al. 1992),

\begin{equation}
\tau\leq3.1\times10^{-8}v_{b}^{3}/n_{x}
\end{equation}
Here, the unit of $v_{b}$ is m s$^{-1}$. Using Equation (2) and
(3) we may explain why the burst lifetime is anti-correlated to
the separate frequency of thee type III pair events.

The plasma temperature can be derived from the ratio of SXR
intensity at the two energy band observed by GOES after
subtracting the background from Solar Software (SSW, Thomas et al.
1985). The right panel of Fig. 3 shows the temporal relation
between the type III pairs and the temperature in the flare
energy-release site. Table 1 lists the temperature in each
microwave type III pair event, which is in a range of (1.06 --
1.85)$\times10^{7}$ K, much hotter than the general coronal plasma
(at the magnitude of million K). However, it is necessary to note
that the above temperature is only related to the hot flaring
loops which is possible different from the real source region of
type III pair bursts. We adopt it just as an approximation to the
source regions of type III pairs.

The frequency drift of microwave type III bursts can be derived
from Equation (1):

\begin{equation}
D=\frac{f}{2n_{e}}(\frac{\partial n_{e}}{\partial
t}+\frac{\partial n_{e}}{\partial l}\cdot \frac{\partial
l}{\partial t})=\frac{f}{2}(\frac{1}{t_{n}}+\frac{v_{b}}{H_{n}})
\end{equation}
And the relative frequency drift is,

\begin{equation}
\bar{D}=\frac{1}{2t_{n}}+\frac{v_{b}}{2H_{n}}
\end{equation}
$H_{n}=\frac{n_{e}}{\partial n_{e}/\partial l}$ is the plasma
density scale length. $H_{n}>0$ or $H_{n}<0$ means the density
increasing or decreasing along the beam propagating direction
($l$). $t_{n}=\frac{n_{e}}{\partial n_{e}/\partial t}$ is the
scale time of plasma density variation. $t_{n}>0$ or $t_{n}<0$
means the density increasing or decreasing with respect to time.
$\frac{\partial l}{\partial t}=v_{b}$ is the velocity of the
electron beam.

Equation (4) and (5) indicate that frequency drift is composed of
two parts. The first part is related to the temporal change of
plasma density ($t_{n}$) around the flare energy-release site due
to plasma flows (such as chromospheric evaporation, reconnecting
flows, and or MHD oscillations). The second part is related to the
plasma density gradient. When the electron beams propagate in the
background plasma it will trigger the Langmuir waves and produce
plasma emission in different frequencies at different locations.

Altyntsev et al. (2007) and Meshalkina et al. (2012) applied the
one-dimensional scan observation at two different frequency
channels, and estimated that the temporal change of plasma density
contributed 4 GHz $s^{-1}$ and 6 GHz $s^{-1}$ to the frequency
drifts in two microwave type III bursts, respectively. However, in
our work we found that most microwave type III pairs have
frequency drifts $<$ 4 GHz $s^{-1}$. Only 3 among the 11 events
have absolute frequency drift $>$ 4 GHz $s^{-1}$, see in Table 1.
Actually, in the framework of plasma emission mechanism the
separate frequency ($f_{x}$) reflects the plasma density in the
flare energy-release site, and the temporal variation of $f_{x}$
may reveal the plasma density change during the microwave type III
burst pair trains. Table 1 listed the temporal change of separate
frequency, most of them are very slow, and even close to zero.
Among the 11 type III pair events, the maximum $\frac{df_{x}}{dt}$
is occurred in the event on 2006 December 13 which is about 10 MHz
s$^{-1}$, and $\frac{df_{x}}{f_{x}dt}\approx 2.9\times10^{-3}$
s$^{-1}$. Then $t_{n}^{-1}=5.8\times10^{-3}$ s$^{-1}$. The
observation shows that the relative frequency drift ($\bar{D}$) is
from 0.3 s$^{-1}$ to 5.3 s$^{-1}$, which is three orders bigger
than that caused by the plasma density temporal variation. It
seems that the plasma density temporal variation does not
contribute to the formation of frequency drift in type III pairs
as much as in microwave type III bursts reported by Altyntsev et
al. (2007) and Meshalkina et al. (2012). The plasma density
gradient dominates the frequency drifts of type III pairs.
Therefore, we have the relation, $\bar{D}\gg t_{n}^{-1}$ and
$\bar{D}\approx\frac{v_{b}}{2H_{n}}$. With these approximations,
the velocity can be expressed,

\begin{equation}
v_{b}\approx2H_{n}\bar{D}.
\end{equation}

The frequency gap ($\delta f$) between the normal and RS branches
is seldom mentioned in previous literature. This work shows that
the observed results of $\delta f$ are ranging from 10\,MHz to
1700\,MHz, and different from event to events. Then, what factors
do dominate the frequency gap? Li et al. (2011) simulated the
generation and propagation of type III bursts based on the
assumption of power-law model (PL, Allen 1947) and offset
power-law model (OPL, Aschwanden \& Benz 1995), respectively. They
find that frequency gaps may be very large ($\delta f>200$\,MHz)
when electron beams propagate in offset power-law plasma, and
relatively small ($\delta f<200$\,MHz) in power-law plasma. It
seems that the frequency gap may be related to the distribution of
plasma density. According to PE mechanism, the frequency gap means
a density difference between start sites of normal and RS type III
bursts. With density scale length $H_{n}$, a space length can be
derived:

\begin{equation}
L_{c}\approx H_{n}\cdot\frac{\delta f}{f_{x}}.
\end{equation}
$L_{c}$ can be regarded as an estimation of the length of
acceleration regions. The electrons are accelerated in this region
and get a relatively high energy, then trigger microwave type III
bursts outside this region. It is possible that $L_{c}$ is only an
upper limit of the acceleration region. Panel (4) of Fig. 5 show
that frequency gap has no dependence on the separate frequency.
This fact implies that the length of acceleration region is
possibly irrelevant to the plasma density. It is possible that the
particle acceleration mechanism is the main factor to dominate the
frequency gap.

Equation (6) and (7) imply that $H_{n}$ is a key factor for
determining the electron beam velocity and the source dimension
length. It depends on the plasma density gradient. The microwave
type III pairs are generated very close to the flare
energy-release site where the magnetic field is an important
factor. Considering magnetic field, free-free absorption (Dulk
1985, Stahli \& Benz 1987), and plasma mechanism, Tan et al.
(2015) proposed a set of formulas to estimate the magnetic field
$B$ and $H_{n}$ near the magnetic reconnection site and the
electron beams energy from the observations of microwave type III
bursts,

\begin{equation}
B_{L}<B<B_{H}
\end{equation}
$B_{L}=3.402\times10^{-19}(n_{x}T\bar{D}R_{c})^{\frac{1}{2}}$ and
$B_{H}=3.293\times10^{-16}[\frac{n_{x}T\bar{D}R_{c}}{(n_{x}\tau)^{\frac{1}{3}}}]^{\frac{1}{2}}$
are the lower and upper limits of the magnetic field,
respectively. As for type III pairs, there are two lower limits
($B_{Ln}$, $B_{Lr}$) and two upper limits ($B_{Hn}$, $B_{Hr}$).
The median of the four values can be the best estimator of
magnetic field near the start site of the electron beams:
\begin{equation}
B_{n}\sim\frac{1}{2}(B_{Ln}+B_{Hn}),
B_{r}\sim\frac{1}{2}(B_{Lr}+B_{Hr}).
\end{equation}
\begin{equation}
H_{n}\approx\frac{\mu_{0}n_{x}k_{B}T}{B^{2}}R_{c}
\end{equation}
\begin{equation}
v_{bn}\approx\frac{2\mu_{0}n_{stn}k_{B}T}{B_{n}^{2}}\bar{D_{n}}R_{c},
v_{br}\approx\frac{2\mu_{0}n_{str}k_{B}T}{B_{r}^{2}}\bar{D_{r}}R_{c}.
\end{equation}
Here, $R_{c}$ is the radius expressing the divergence of magnetic
field lines. Applying this method to the type III pair trains of
2006 December 13, we found that the magnetic field is from
53.6$\pm$17.0 G to 87.4$\pm$27.2 G, the energy of the downward
electron beams is 42 - 64 keV which is much similar to that of the
upward electron beams. The length of acceleration region is about
18 - 733 km. As lack of imaging observations and $R_{c}$ is
unknown, we did not make the similar estimations in other type III
pair events listed in Table 1 and 2.

\section{Conclusions}

This work reports 11 microwave type III pair events observed by
two radio spectrometers at frequency range of 0.80 - 7.60\,GHz in
9 solar flares during 1994 - 2014. In the following, we present
our conclusions which may provide new constraints in understanding
of the origin of solar eruptions and the energetic particles:

(1) All microwave type III pairs occurred after flare onset,
distributed in flare impulsive and postflare phases. No type III
pair is observed in preflare phase so far. This fact is different
from the existing model (Aschwanden \& Benz 1995).

(2) The parameters of microwave type III pairs are distributed in
a wide ranges, e.g. the separate frequency is 1.08 - 3.42\,GHz,
the frequency gap is 10 - 1700\,MHz, and frequency drifts is 0.4 -
14.5\,GHz\,s$^{-1}$. The temporal change of separate frequency is
very small, and the separate frequency is near a constant in each
type III pair train. The most peculiar is that, in most cases, the
normal type III branches drift obviously faster than the RS
branches. These parameters are different from previous literature
(Aschwanden et al. 1997, Ning et al. 2000, etc.). These facts
imply that flare primary energy release and electron acceleration
may take place in a wider space range.

(3) There are obvious differences between type III pairs occurring
in flare impulsive phase and those occurring in postflare phase.
Among the 11 type III pair events, 7 events occurred in flare
impulsive phases and the other 4 events occurred in postflare
phases. The former have relatively low separate frequency, long
lifetime, large frequency gap, and slower frequency drift. The
latter have relatively higher separate frequency, shorter
lifetime, narrower frequency gap, and faster frequency drift, and
strong circular polarization.

(4) The frequency drift increases with the separate frequency, the
lifetime of each individual burst is anti-correlated to the
separate frequency, while the frequency gap between the normal and
RS branches of type III pair seems to be independent to the
separate frequency.

(5) The plasma density around the flare energy-release sites is
about $10^{10} - 10^{11}$\,cm$^{-3}$. The SXR observations show
the temperature near source regions are about $(1.08 -
1.85)\times10^{7}$ K, much hotter than the quiet coronal plasma.

Using the new method of Tan et al. (2015) we may estimate the
magnetic field and plasma density near the acceleration regions
and energy of the electron beams from the observation of microwave
type III pairs. Here the imaging observation at the corresponding
frequency is necessary. This work demonstrates that broadband
spectral imaging observations at microwave range are crucial for
understanding the primary energy release and particle
accelerations in solar flares. Combining with studies of other
spectral fine structures (Huang et al., 2007, Altyntsev et al.
2007, etc.) and imaging observations (Yan et al. 2009), the
intrinsic natures of solar flares can be revealed.

\acknowledgments The authors thank Profs A. Altyntsev, H. Hudson,
and V. Melnikov, Drs L. Kashapova and E. Kontar for helpful
suggestions and valuable discussions on this paper. We would also
thank the ORSC/Ond\v{r}ejov and SBRS/Huairou teams for providing
observation data. B.T. and C.T. acknowledges support by the NSFC
Grants 11273030, 11221063, 11373039, 11433006, and 2014FY120300,
CAS XDB09000000, R\&D Project ZDYZ2009-3. H.M. and M.K.
acknowledges support by the Grant P209/12/00103 (GA CR) and
Project RVO: 67985815 of the Astronomical Institute AS. This work
is also supported by the Marie Curie PIRSES-GA-295272-RADIOSUN
project.


\begin{thebibliography}{}

\bibitem[Allen(1947)]{Allen1947}Allen, C. W.: 1947, MNRAS, 107, 426

\bibitem[Altyntsev(2007)]{Altyntsev2007}Altyntsev, A. T., Grechnev, V. V., \& Meshalkna, N. S.: 2007, SoPh, 242, 111

\bibitem[Aschwanden(1993)]{Aschwanden1993}Aschwanden, M. J., Benz, A. O., \& Schwartz, R.A.: 1993, ApJ, 417, 790

\bibitem[Aschwanden(1995)]{Aschwanden1995}Aschwanden, M. J., \& Benz, A. O.: 1995, ApJ, 438, 997

\bibitem[Aschwanden(1997)]{Aschwanden1997}Aschwanden, M. J., \& Benz, A. O.: 1997, ApJ, 480, 825

\bibitem[Bastian(1998)]{Bastian1998}Bastian, T. S., Benz, A. O., \& Gary, D. E.: 1998, ARA$\&$A, 36, 131

\bibitem[Benz(1992)]{Benz1992}Benz, A. O., Magun, A., Stehling, W., \& Su, H: 1992, SoPh, 141, 335

\bibitem[Chen(2008)]{Chen2008}Chen, B., \& Yan, Y. H.: 2008, ApJ, 689, 1412

\bibitem[Chen(2013)]{Chen2013}Chen, B., Bastian, T.S., White, S.M., \& et al.: 2013, ApJL, 763, 21

\bibitem[Christe(2008)]{Christe2008}Christe, S., Krucker, S., Lin, R. P.: 2008, ApJ, 680, 149

\bibitem[Dulk(1985)]{Dulk1985}Dulk, G. A.: 1985, ARA$\&$A, 23, 169

\bibitem[Fu(1995)]{Fu1995} Fu, Q. J., Qin, Z. H., Ji, H. R., \& et al: 1995, SoPh, 160, 97

\bibitem[Fu(2004)]{Fu2004} Fu, Q. J., Ji, H. R., Qin, Z. H. \& et al.: 2004, SoPh, 222, 167

\bibitem[Huang(1998)]{Huang1998} Huang, G. L., Qin, Z. H., Yan, G., Fu, Q. J., \& Liu, Y. Y.: 1998, Ap$\&$SS, 259, 317

\bibitem[Huang(2007)]{Huang2007} Huang, J., Yan, Y.H, \& Liu, Y.Y.: 2007, Adv Space Res., 39, 1439

\bibitem[Huang(2011)]{Huang2011} Huang, J., Demoulin, P., Pick, M., \& et al.: 2011, ApJ, 729, 107

\bibitem[Jiricka(1993)]{Jiricka1993} Ji\v{r}i\v{c}ka, K., Karlick\'y, M., Kepka, O., \& Tlamicha, A.: 1993, SoPh, 147, 203

\bibitem[Kane(1981)]{Kane1981} Kane, S. R.: 1981, ApJ, 247, 1113

\bibitem[Karlicky(2014)]{Karlicky2014} Karlick\'y, M.: 2014, RAA, 14,
753

\bibitem[Kliem(2000)]{Kliem2000}Kliem, B., Karlick\'y, M., \& Benz, A. O.: 2000, A$\&$A, 360, 715

\bibitem[Li(2011)]{Li2011} Li, B., Cairns, I.H., Yan, Y.H., \& Robinson, P.A.: 2011, ApJL, 738, 9

\bibitem[Lin(1971)]{Lin1971}Lin, R. P., \& Hudson, H. S.: 1971, SoPh, 17, 412

\bibitem[Lin(1981)]{Lin1981}Lin, R. P., Potter, D. W., Gurnett, D. A., \& Scarf, F. L.: 1981, ApJ, 251,
364

\bibitem[Ma(2008)]{Ma2008} Ma, Y., Wang, D. Y., Xie, R. X., Wang, M., \& Yan, Y. H.: 2008, Ap$\&$SS, 318, 87

\bibitem[Masuda(1994)]{Masuda1994} Masuda, S., Kosugi, T., Hara, H., Tsuneta, S., \& Ogawara, Y.: 1994, Nature, 371,
495

\bibitem[Meszarosova(2008)]{Meszarosova2008}M\'esz\'arosov\'a, H., Karlick\'y, M., Sawant, H.S., et al.: 2008, A$\&$A, 484,
529

\bibitem[Melrose(1982)]{Melrose1982} Melrose, D.B., \& Dulk, G.A.: 1982, ApJ, 259, 844

\bibitem[Meshalkina(2004)]{Meshalkina2004}Meshalkina, N. S., Altyntsev, A.T., Sych, R.A., Chernov, G.P., \& Yan, Y.H.: 2004, SoPh, 221,
85

\bibitem[Meshalkina(2012)]{Meshalkina2012}Meshalkina, N. S., Altyntsev, A.T., Zhdanov, D.A., \& et al: 2012, SoPh, 280, 537

\bibitem[Ning(2000)]{Ning2000}Ning, Z. J., Fu, Q. J., \& Lu, Q. K.: 2000, SoPh, 194, 137

\bibitem[Reid(2014)]{Reid2014} Reid, H. A. S., \& Ratcliffe, H.: 2014, RAA, 14, 773

\bibitem[Robinson(2000)]{Robinson2000}Robinson, P. A., \& Benz, A. O.: 2000, SoPh, 194, 345

\bibitem[Sakai(2005)]{Sakai2005}Sakai, J. I., Kitamoto, T., Saito, S.: 2005, ApJ, 622, 157

\bibitem[Sawant(1994)]{Sawant1994}Sawant, H. S., Fernandes, F. C., \& Neri, J. A. C.F.: 1994, ApJS, 90, 689

\bibitem[Stahli(1987)]{Stahli1987}Stahli, M., Benz, A. O.: 1987, A$\&$A, 175, 271

\bibitem[Tan(2007)]{Tan2007}Tan, B. L., Yan, Y.H., Tan, C. M., \& Liu, Y.Y.: 2007, ApJ, 671, 964

\bibitem[Tan(2010)]{Tan2010}Tan, B. L., Zhang, Y., Tan, C. M., \& Liu, Y.Y.: 2010, ApJ, 723, 25

\bibitem[Tan(2013)]{Tan2013}Tan, B. L.: 2013, ApJ, 773, 165

\bibitem[Tan(2015)]{Tan2015}Tan, B. L., Karlick\'y, M., M\'esz\'arosov\'a, H., \& Huang, G.L.: 2015, RAA,
accepted

\bibitem[Tang(2012)]{Tang2012}Tang, J. F., Wu, D.J., \& Yan, Y. H.: 2012, ApJ, 745, 134

\bibitem[Thomas(1985)]{Thomas85} Thomas R.J., Starr R., \& Crannell C.J., 1985, SoPh, 95, 323

\bibitem[Yan et al(2002)]{Yan2002} Yan, Y.H., Tan, C.M., \& Xu, L., et al., 2002, Sci. Chin. A Suppl., 45, 89

\bibitem[Yan et al.(2009)]{Yan2009} Yan, Y. H., Zhang, J., \& Wang, W., et al.: 2009, EM$\&$P, 104, 97

\bibitem[Zheleznyakov(1975)]{Zheleznyakov75} Zheleznyakov, V.V., \& Zlotnik, E.YA.: 1975, SoPh, 44, 461

\end{thebibliography}
\end{document}